\documentclass[aps,pra,superscriptaddress,a4paper,reprint]{revtex4-1}

\usepackage{amsmath}
\usepackage{graphicx}
\usepackage[left=2cm,right=2cm,
    top=2.5cm,bottom=2.5cm,bindingoffset=0cm]{geometry}

\usepackage{hyperref}
\hypersetup
{%
  pdftitle = {Shifting the phase of a coherent beam with a Yb+ ion},
  pdfauthor = {Markus Sondermann},
  colorlinks = {false}
}

\begin{document}
\title{Shifting the phase of a coherent beam with a $^{174}$Yb$^+$ ion:\\
  influence of the scattering cross section}

\author{Martin Fischer}
\thanks{These authors contributed equally to this work.}
\affiliation{Max Planck Institute for the Science of Light,
  Guenther-Scharowsky-Str. 1/ building 24, 91058 Erlangen, Germany}
\email{gerd.leuchs@mpl.mpg.de}
\affiliation{Friedrich-Alexander-Universit\"at Erlangen-N\"urnberg (FAU),
  Department of Physics, Staudtstr. 7/B2, 91058 Erlangen, Germany}

\author{Bharath Srivathsan}
\thanks{These authors contributed equally to this work.}
\affiliation{Max Planck Institute for the Science of Light,
  Guenther-Scharowsky-Str. 1/ building 24, 91058 Erlangen, Germany}

\author{Lucas Alber}
\thanks{These authors contributed equally to this work.}
\affiliation{Max Planck Institute for the Science of Light,
  Guenther-Scharowsky-Str. 1/ building 24, 91058 Erlangen, Germany}
\affiliation{Friedrich-Alexander-Universit\"at Erlangen-N\"urnberg (FAU),
  Department of Physics, Staudtstr. 7/B2, 91058 Erlangen, Germany}

\author{Markus Weber}
\affiliation{Max Planck Institute for the Science of Light,
  Guenther-Scharowsky-Str. 1/ building 24, 91058 Erlangen, Germany}

\author{Markus Sondermann}
\affiliation{Max Planck Institute for the Science of Light,
  Guenther-Scharowsky-Str. 1/ building 24, 91058 Erlangen, Germany}
\affiliation{Friedrich-Alexander-Universit\"at Erlangen-N\"urnberg (FAU),
  Department of Physics, Staudtstr. 7/B2, 91058 Erlangen, Germany}

\author{Gerd Leuchs}
\affiliation{Max Planck Institute for the Science of Light,
  Guenther-Scharowsky-Str. 1/ building 24, 91058 Erlangen, Germany}
\affiliation{Friedrich-Alexander-Universit\"at Erlangen-N\"urnberg (FAU),
  Department of Physics, Staudtstr. 7/B2, 91058 Erlangen, Germany}
\affiliation{Department of Physics, University of Ottawa, Ottawa,
  Ont. K1N 6N5, Canada}

\begin{abstract}
 We discuss and measure the phase shift imposed onto a radially
 polarized light beam  when focusing it onto an $^{174}\text{Yb}^{+}$
 ion.
 In the derivation of the expected phase shifts we 
 include the properties of the involved atomic levels.
 Furthermore, we emphasize the importance of the
 scattering cross section and its relation to the efficiency for coupling
 the focused light to an atom.
 The phase shifts found in the experiment are compatible with the
 expected ones when accounting for known
 deficiencies of the focusing optics and the motion of the trapped ion
 at the Doppler limit of laser cooling\,\cite{haensch1975}. 
\end{abstract} 

\maketitle

\section{Recollection by one of us (GL)}
{\it
Birthdays often evoke memories of the one who is
celebrating. Sometimes it is a single question they have asked you
that has stuck in your mind for a long time. Of the many times I met
Ted H\"ansch one comes to my mind in particular. It was when I saw him
in a corridor at the Max Planck Institute of Quantum Optics, about
thirty years ago --  the building was quite new at the time. I
vividly remember the question he asked me: \lq Do you have a good
explanation why the cross section of an atom for scattering light
is as large as it is?\rq\   
He was referring to the classical on-resonance cross section of an atom,
$\sigma_{\text{sc}}=3\lambda^2/2\pi$, being so much larger –- i.e. many
orders of magnitude –- than the cross section of the atomic charge
distribution. Naturally, I knew the phenomenon and answered that in
scattering processes the larger of the two following values dominates:
the cross section of the atom as a massive object or the cross section
of the particle you send in to probe the atom, namely a photon in the
case under consideration. Obviously the smallest cross section of an
optical beam is limited by diffraction and this, I had thought, should
define the cross section of the photon. I was surprised to see that
Ted H\"ansch did not seem satisfied as he slowly turned away. At the
time this made me think, and throughout the years since then I have
returned to this thought every now and then.

Ten years later, after I moved to Erlangen, this \lq thinking\rq\
became more intense when within my group we started to first discuss
spontaneous  
emission and the possibility of observing its time reversed
counterpart. In spontaneous emission the energy is initially
concentrated in a tiny volume, which is orders of magnitude smaller
than the wavelength cubed -- partially still stored in the atom -- and
begins to travel outwards. At first the energy is both in the
evanescent and propagating components of the field. Then, as the
outgoing dipole wave travels further, the evanescent components decay
away leaving only the propagating part of the dipole wave. The idea
arose that the evanescent field is more part of the atom than of the
out-going dipole wave.
Consulting any book on electromagnetism, one can
calculate the outward going energy flux of the near field and that of
the propagating field. The near field part quickly decreases to zero
as the distance to the source increases, whereas the far field
component is constant. It is interesting to note that the radial
position $r$, at which the energy flux of the near field has reduced
to half the far field portion, is given by
$(2\pi/\lambda)^2r^2\sim6$. This $r$ value corresponds exactly to the
above mentioned cross section, indicating that in terms of cross
section the near field can be considered part of the atom.
Was this good enough to tell Ted H\"ansch? Without the
atom, light would produce a diffraction-limited spot, but when an atom
is at the origin of the dipole wave one instead expects time reversed
spontaneous emission to occur such that the energy density of the
field increases far beyond the diffraction limited value in free
space. One might speculate that the evanescent field is excited via
the atom's reaction to the incident field. If one considers the
evanescent field as part of the atom its extent defines the cross
section of the atom, resulting in a cross section almost matching the
classical textbook value quoted above. Nevertheless, at that point I
still felt it was too early to go back to Ted H\"ansch. There was still
something that puzzled me.

The incoming dipole wave with its evanescent and propagating
components is an exact solution of Maxwell's equations but it has a
singularity. Accordingly, when one excites an inward propagating
dipole wave in the far field one would expect the singularity to
develop -- this is part of the rigorous solution after all -- up to
the point when the wave reaches the atomic charge distribution. We
know however that this is not what happens. Thus, it was
a great relief to me when Simon Heugel, a doctoral student in our
group at the time, came to me about seven years ago suggesting that I
look at problem C$_I$.6 in the text book by Cohen-Tannoudji, Dupont-Roc and
Grynberg\,\cite{cohen-tannoudji1989}.
There it is stated that in free space the inward propagating dipole
will continue as an outward propagating dipole once it has passed the
origin and will thus interfere with itself. The task given to the
students is to calculate the energy density of the resulting standing
wave and -- alas -- the result is the diffraction limited field
distribution, provided one takes into account a phase shift at the
origin which is in a way the Gouy phase shift under this extreme full
solid angle focusing scenario. Looking at the problem in this way
everything seems to fall into place: (1) when focusing in free space
the singular terms in the dipole wave solution interfere
destructively, and (2) suppressing the outward going wave via full
absorption at the origin by a sub-wavelength antenna such as an atom
gives rise to the well known field enhancement. We asked
ourselves whether there are other ways to restore the singular
behavior. One way we found theoretically was by studying the time
evolution of the energy distribution for focusing in free space near
the origin when the inward going dipole wave has a sharp rising
leading edge, rising over a distance significantly smaller than the
wavelength. This indeed also gave a transient
enhancement\,\cite{gonoskov2012}. Other experiments are under way. 

Encouraged by these considerations and findings we hope this
anniversary is the right moment to give Ted H\"ansch an update on our,
by now decades long, attempt to answer the question he posed such a
long time ago. 
}

\section{Introduction}
\label{intro}

The \emph{scattering cross section} is a quantity used in many areas
in physics,
relating the rate of particles scattered by a target to the flux of
particles incident onto it.
In quantum optics, the conceptually simplest target is a single atom
and the incident particles are photons.
For this scenario, the resonant scattering cross section for a
two-level atom is determined to be\,\cite{mandel-wolf1995, budker2004} 
\begin{equation}
\label{eq:sigma}
\sigma_{\text{sc}} = \frac{3}{2\pi}\lambda^2
\end{equation}
for an atomic transition with resonance wavelength $\lambda$
provided the oscillator strength\,\cite{sobelman1972} is equal to one.

As mentioned above, the area given by $\sigma_{\text{sc}}$ is large:
It is by far larger than the spatial extent of an atom given
by the Bohr radius and also larger than the smallest spot sizes
achievable via diffraction limited focusing of light with lenses of
sufficient numerical aperture\,\cite{Quabis_OptComm2000, dorn2003}.
The term cross section was created to describe scattering of
particles, but in wave mechanics there is also the interference of fields.
As pointed out in Ref.\,\cite{cray1982} absorption can be described as
the interference of the (non attenuated) incident field and the
scattered field.
In this model attenuation in forward direction is caused by the
destructive interference between these two fields, which requires a
power of the scattered field which 
may seem counter-intuitive at first sight: full attenuation and only
back scattered light requires the power of the scattered field to be
twice that of the incident field because of the destructive interference with the
incident light in the forward direction, in order to fulfill energy conservation.
Along those lines the rate of scattered photons, which is not to be
confused with the detected photons,
expressed in terms of cross sections is given by\,\cite{zumofen2008}

\begin{equation}
  \label{eq:ratio}
  \gamma_{\text{sc}}=\frac{\sigma_{\text{sc}}}{A}\cdot \gamma_{\text{inc}},
\end{equation}

with $A$ denoting the effective mode area\,\cite{zumofen2008, domokos2002} of the
incident stream of photons $\gamma_{\text{inc}}$. 
The remarkable scenario of more photons being 
scattered than photons arriving, both per unit time \cite{kochan1994}, arises when
$\sigma_{\text{sc}}$ becomes larger than $A$.
Due to the interference of the different outward going partial
waves energy conservation is, of course, maintained.
Within this reasoning, several intriguing phenomena occurring in the
interaction of light and single quantum emitters have been
investigated in recent years, see Ref.\,\cite{leuchs2013o} for a
review.
However, as reported in Ref.\,\cite{Tretyakov2014} it was found
already in the early 1980's by
\emph{Bohren}\,\cite{bohren1983} and \emph{Paul \& Fischer}\,\cite{paul1983} 
that an atom can scatter more light than incident onto its massive
cross section, which is on the order of the Bohr radius squared.
As also discussed in more recent publications, the key step
in these papers was indeed the examination of the superposition of incident
and scattered fields.
Refs.\,\cite{bohren1983, paul1983} revealed that within a certain area
larger than the size of the scatterer the resulting lines of energy
flux end up at the scatterer's position.
Within a similar reasoning and as outlined in the first section of this
paper, one could attribute the spatial extent of the
non-propagating near-field components of the field re-radiated by the
atom to the size of the atom, leading to the expression for
$\sigma_{\text{sc}}$ given by Eq.\,\ref{eq:sigma}.  

Here, we relate to such concepts by investigating the phase shift
imprinted onto a tightly focused light beam by a single
$^{174}\text{Yb}^{+}$ ion.
In the next section, the importance of the magnitude of the effective
mode area of the incident beam to the obtained phase shift is
revisited.
With simple arguments, we modify the equation obtained in
Refs.\,\cite{sondermann2013p, leuchs2013o} describing the
achievable phase shift to account for the level structure of the used
ion species.
Explicitly, we make use of the dependence of the scattering cross
section on the angular momenta of the involved atomic levels.
In Sec.\,\ref{sec:exp} we describe our experimental apparatus, present
the phase shift observed in our experiments and compare the obtained
results to the predictions of Sec.\,\ref{sec:frame}.
At the end of the paper we give concluding remarks.

\section{Relation of scattering cross section and
  phase shift}
\label{sec:frame}

In order to emphasize the role  of the scattering cross section
$\sigma_{\text{sc}}$ in phase shifting a weak coherent beam we briefly
recall some essential aspects.
Typically, the induced phase shift is considered as the phase
difference of the superposition of the incident electric field
$E_{\text{inc}}$ and the scattered field $E_{\text{sc}}$ relative to
the phase of the incident one, i.e. the phase of the incident field
leaving the interaction region when no atom is
present\,\cite{PhysRevLett_kurtsiefer, Sandoghdar_phase, Blatt_phase}. 
The phase shift $\Delta\varphi$ can then be written
as\,\cite{PhysRevLett_kurtsiefer}
\begin{equation}
\Delta \varphi=\arg\left(\frac{E_{\text{inc}}+E_{\text{sc}}}{E_{\text{inc}}}\right)
\end{equation}
with $\arg(\,)$ denoting the argument of its complex variable.

Since one is considering a coherent process in this situation it is detrimental to
saturate the atomic transition, i.e. to produce incoherent components
in the scattered radiation.
We therefore assume negligible saturation.
For this case the phase shift imprinted by a
\emph{two-level} atom is found to be\,\cite{sondermann2013p}
\begin{equation}
  \label{eq:phaseTL}
  \Delta\varphi=\arg\left(1-2G\cdot\frac{1+i\cdot2\Delta/\Gamma}
                         {1+4\Delta^2/\Gamma^2}\right)\, . 
\end{equation}
Here, $\Gamma$ denotes the spontaneous emission rate and 
$\Delta$ is the detuning between the laser and the
atomic resonance frequency.
At fixed detuning, the crucial parameter determining the magnitude of
the imprinted phase shift is $G$, describing the extent to which the
atom experiences the highest possible electric field at constant input
power which is allowed for by diffraction: $G=
E_{\text{inc}}^2/E_{\text{max}}^2$, where $0\le G\le1$.
$E_{\text{max}}$ is the field amplitude obtained by focusing a dipole wave 
in free space\,\cite{bassett_limit_1986}, i.e.
$G$ determines how efficiently the
incident field couples to the atomic dipole transition.
Assuming an atom at rest, $G$ is solely determined by the properties
of the focusing optics and the spatial mode of the incident
field which has an overlap of $\eta$ with the field of the driven
transition\,\cite{leuchs2013o, golla2012}, $G\propto\eta^2$.
It also accounts for phase front aberrations that are induced 
by imperfect focusing optics\,\cite{fischer_efficient_2014, alber2016}.  
Therefore, $G$ is a measure for the quality of the mode matching
of the incident mode to the atomic dipole-radiation pattern.

The role of $G$ becomes obvious when relating it to the so-called
\emph{scattering ratio} on resonance, which is defined as
$R=\gamma_{\text{sc}}/\gamma_{\text{inc}}$\,\cite{zumofen2008, PhysRevLett_kurtsiefer}.
One can show that in general $G=R/4$\,\cite{leuchs2013o}, resulting in
\begin{equation}
  \label{eq:Gsigma}
  G=\frac{\sigma_{\text{sc}}}{4A} \, .
\end{equation}
Hence, in order to reach unit coupling efficiency and thus
the maximum phase shift at a fixed, non-zero detuning, the
effective mode area of the focused beam must not be larger than a quarter of the
scattering cross section. One can actually show that
$\sigma_{\text{sc}}/4$ is the minimum possible mode area in free
space\,\cite{minA_note}.
This minimum is obtained at
$G=1$\,\cite{leuchs2013o}.

Inserting Eq.\,\ref{eq:Gsigma} into Eq.\,\ref{eq:phaseTL} results in
\begin{equation}
  \label{eq:phaseSigma}
  \Delta\varphi=\arg\left(1-\frac{\sigma_{\text{sc}}}{2A}
  \cdot\frac{1+i\cdot2\Delta/\Gamma}
                         {1+4\Delta^2/\Gamma^2}\right)\, , 
\end{equation}
similar to the findings of Ref.\,\cite{PhysRevLett_kurtsiefer}.
On resonance, the phase of the outgoing field can only take two
values: zero if $A\geq\sigma_{\textrm{sc}}/2$ and $\pi$ as soon as the
electric field is focused to a spot smaller than
$\sigma_{\text{sc}}/2$.
This representation reveals that the obtainable phase shift is not only
limited by imperfect focusing, as expressed by a too large $A$.
But also choosing the \lq ideal\rq\  atom is of importance, i.e. an atom
for which Eq.\,\ref{eq:sigma} is valid.
Deviations could originate from a degenerate ground state as is the
case for $^{174}\text{Yb}^+$ or from an atom not being at
rest\,\cite{budker2004}.
Both obstacles occur in the experiment presented in the 
next section.

In the remainder of this section we explicitly treat the level structure.
In general, when accounting for the sub-structure of the atomic levels
involved, the resonant scattering cross
section of an atomic transition can be written as\,\cite{budker2004}
\begin{equation}
  \sigma_{\text{sc}}=\frac{\lambda^2}{2\pi}\cdot\frac{2J'+1}{2J+1}
\end{equation}
with $J'$ and $J$ being the total angular momentum of upper and lower
level, respectively.
For our experiment involving the $\text{P}_{1/2}\leftrightarrow\text{S}_{1/2}$
transition of $^{174}\text{Yb}^+$ (cf. Fig.\,\ref{fig:levels}) we have $J'=J$
and hence $\sigma_{\text{sc}}=\lambda^2/(2\pi)$, which is only $1/3$
of the value used so far.
We explicitly account for this reduction of the scattering cross
section in writing
\begin{equation}
  \label{eq:phaseJJ}
  \Delta\varphi_{J=J'}=\arg\left(1-\frac{2G}{3}\cdot\frac{1+i\cdot2\Delta/\Gamma}
                         {1+4\Delta^2/\Gamma^2}\right)\, . 
\end{equation}
Consequently, $G$ from now on only accounts for imperfect focusing and
atomic motion. 

The result of Eq.\,\ref{eq:phaseJJ} can also be obtained from a
solution of the Bloch equations for a $J'=1/2\leftrightarrow J=1/2$
system driven only by a $\pi$-polarized light field.
The modification $G\rightarrow G/3$ can be understood as
follows. 
First, the dipole moment in excitation is reduced
by a factor $1/\sqrt{3}$ in comparison to a two-level atom.
Second, the amplitude of the coherently scattered field that
can interfere with the incident radiation, is reduced by the
same factor, because the $\sigma^\pm$-components of the scattered field
cannot interfere with the incident light.
A detailed calculation will be presented somewhere else.

\section{Set-up and Experiment}
\label{sec:exp}

\begin{figure}
\centering
\resizebox{1.0\columnwidth}{!}{\includegraphics{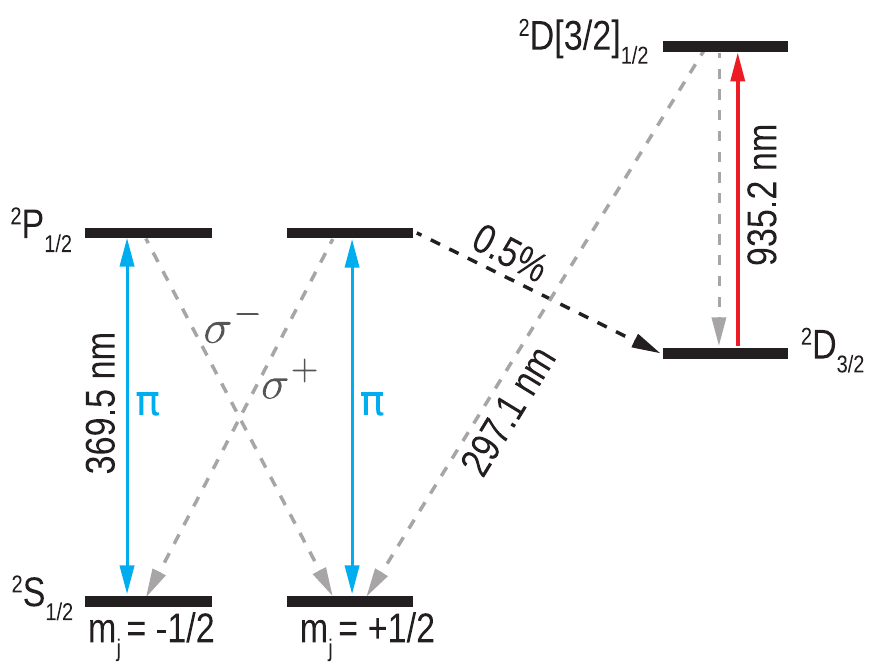}}
\caption{\label{fig:levels}
  Level scheme of $^{174}\text{Yb}^{+}$. In the phase-shift experiments,
  we drive the $\pi$-transition between the
  $S_{\text{1/2}}$ and the $P_{\text{1/2}}$ state. Furthermore, we use
  optical pumping to prepare the ion in the
  metastable $D_{\text{3/2}}$ (dark) state for obtaining a reference
  phase.
  The branching ratio from
  the $P_{\text{1/2}}$ state into the $D_{\text{3/2}}$ state is
  $0.5\,\%$ \cite{olmschenk_manipulation_2007}. } 
\end{figure}

\begin{figure}

\resizebox{1.00\columnwidth}{!}{\includegraphics{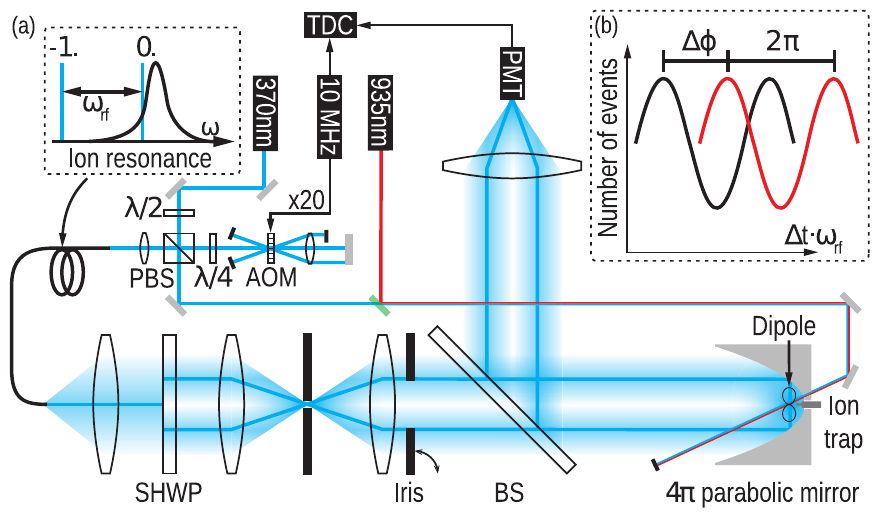}}
\caption{\label{fig:setup}
  (a) Set-up for measuring the phase shift imprinted by a single
  $^{174}\text{Yb}^+$-ion. With an additional laser beam at a
  wavelength of 935.2\,nm we can pump the ion back from the metastable
  $D_{\text{3/2}}$ state into the $S_{\text{1/2}}$ ground state. This
  laser is sent onto the ion from a hole at the backside of the
  parabolic mirror. The same is done for a second laser beam at a
  wavelength of 369.5\,nm that is used for ionization and
  for cooling the ion in certain steps of the experimental procedure
  (see text).
  SHWP: Segmented half-wave-plate, (P)BS: (polarizing) beam splitter
  (other abbreviations in the text). (b) Intensity
  signal $I_{\text{TDC}}(∆t)$ obtained from the statistics of photon
  detection times on the TDC for the ion being in the bright state
  (red) and in the $D_{\text{3/2}}$ dark state (black). 
}
\end{figure}

In our set-up we utilize a parabolic mirror as
the focusing device\,\cite{maiwald_collecting_2012, fischer_efficient_2014, alber2016}.
The parabolic mirror tightly
focuses a radially polarized donut mode to a field that is linearly
polarized along the optical axis
\cite{Quabis_OptComm2000, sondermann_design_2007}.
This field drives a linear dipole oriented in the same
direction.

We position the $^{174}\text{Yb}^+$ ion in the focus of the
mirror by means of a movable open-access ion
trap\,\cite{maiwald_collecting_2012}.
The focused donut mode continuously drives the linear dipole of
the $\text{S}_{1/2}$ to $\text{P}_{1/2}$ transition of the ion, with
a linewidth of
$\Gamma/2\pi=19.6\,\textrm{MHz}$\,\cite{olmschenk_manipulation_2007},
at a wavelength of 369.5\,nm.
The power of this beam is chosen such that saturation effects are
negligible. 
In Ref.\,\cite{alber2016} it was found that the aberrations of
the parabolic mirror used are so strong in the outer parts
that it is favorable to focus only from 
the \lq backward\rq\ half space when not correcting for these aberrations.
We therefore decided to use this focusing configuration in the
experiment reported here, by inserting a suitable iris in the excitation beam
path, cf. Fig.\,\ref{fig:setup}.
The iris has a radius of two times the focal length of the
paraboloid. We refer to this configuration as focusing from half
solid angle, since the bore in the vertex of the parabolic mirror,
housing the trap, reduces the solid angle, relevant for the linear
dipole, by less than 0.5\%.
The focused donut mode also provides Doppler cooling for the ion.
Auxiliary beams needed e.g. for the initial ionization and trapping as
well as the repumping beam (cf. Figs.\,\ref{fig:levels}
and\,\ref{fig:setup}) are entering the focal region of
the mirror through a small auxiliary hole close to the vertex of the
parabola.

Each phase shift measurement is preceded by the following sequence:
First the ion is Doppler cooled by an auxiliary beam red detuned by
half a linewidth from the
$\text{S}_{1/2}\leftrightarrow\text{P}_{1/2}$-transition.
Then this auxiliary beam is switched off and the donut mode drives the
ion at half linewidth detuning.
The ion is scanned through the focal region while
monitoring the count rate of photons at 297\,nm, see
Fig.\,\ref{fig:levels}.
The ion is positioned such that this count rate is maximized.
Afterwards, the auxiliary beam at 369.5\,nm is switched on again for Doppler
cooling.
Switching this beam off again and setting the donut beam to the
desired detuning, the phase shift measurement is initiated.

In this measurement interval, the temperature of the ion is governed
by the interaction with the donut beam.
Hence, the temperature is explicitly depending on the detuning of the
donut beam, as also discussed later.
For a detuning of $\Delta=-\Gamma/2$ Doppler cooling
theory\,\cite{stenholm1986} predicts a
minimal temperature of the ion of about
$\textrm{T}=\hbar\Gamma/2\textrm{k}_{\textrm{B}}=470\,\mu\textrm{K}$, 
where $\hbar$ is Planck's constant and $\textrm{k}_{\textrm{B}}$ the
Boltzmann constant.
From experimentally measured point spread functions (see
Ref.\,\cite{alber2016}) and the characteristics of our ion trap (trap
frequencies of 480\,kHz and 1025\,kHz in radial and axial direction,
respectively), we determine an upper bound of the ion's temperature
to be 50\% above the Doppler limit at half linewidth detuning.

The phase shift imprinted by the ion is measured in a common path
interferometer by heterodyne detection.
We illuminate the ion with the near-resonant carrier donut mode and an
off-resonant sideband donut, similar to the technique applied in
Ref.\,\cite{Sandoghdar_phase}.
The sideband donut is red-detuned from the
$\text{S}_{1/2}\leftrightarrow\text{P}_{1/2}$-transition
by $\omega_{\text{rf}}/2\pi=400$\,MHz (amounting to about 20
linewidths) by using the diffraction order \lq -1\rq\   of an 
acousto-optical modulator (AOM) in double pass configuration
($\omega_{\text{rf}} = 2\, \omega_{\text{AOM}}$, see
Fig. \ref{fig:setup}).
Except for the
frequency difference, both beams have exactly the same properties and
are in the same spatial mode that is focused onto the ion. 

After interaction with the trapped ion, the beams are retro reflected
and recollimated by the parabolic mirror.
We measure the beating signal of the two beams with a
correlation measurement (Fig.\,\ref{fig:setup}) involving a photomultiplier
tube (PMT), a time to digital converter (TDC), and a $10$\,MHz trigger
signal that is synchronized to the AOM, respectively. The intensity
signal $I_{\text{TDC}}(\Delta t)$ obtained from the statistics of
photon detection times on the TDC is fitted with a function proportional
to $\cos(\omega_{\text{rf}}\, \Delta t + \phi_1)$ with phase offset $\phi_1$.
To infer the relative phase shift $\Delta\varphi$ of the
near-resonant beam, we repeat the measurement and fitting procedure
after preparing the ion in the metastable $\text{D}_{3/2}$ (dark) state
by optical pumping (see Fig. \ref{fig:levels}).
This second measurement delivers the reference phase offset $\phi_2$.
The phase shift $\Delta\varphi$ of the near resonant beam is finally
calculated via $\Delta\varphi=\phi_1-\phi_2$.
The acquisition of sufficient statistics for each data point
takes approximately ten seconds.

\begin{figure}
\centering
\resizebox{1.00\columnwidth}{!}{\includegraphics{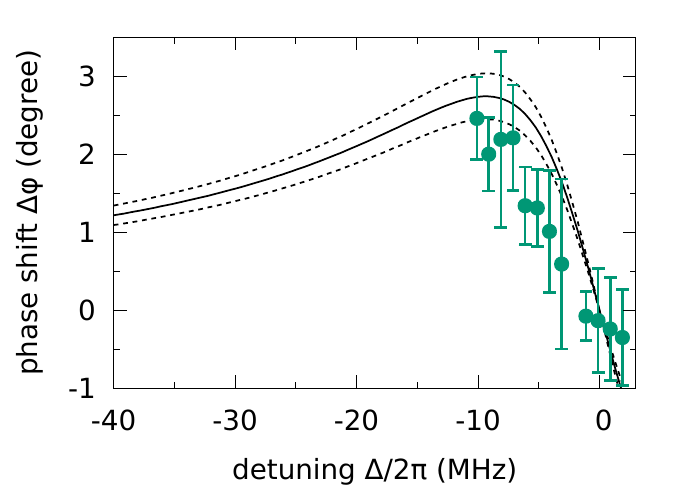}}
\caption{\label{fig:results}
  Measured phase shift $\Delta \varphi$ for different detunings
  (symbols) and phase shift according to Eq. \ref{eq:phaseJJ} for a coupling
  efficiency of $G=13.7 \pm 1.4\,\%$ (solid and dashed lines).
The value used for $G$ is the one found in a saturation measurement in
Ref.\,\cite{alber2016}.}    
\end{figure}

The results for measuring the phase shift as a function of detuning
are shown in Fig. \ref{fig:results}. We achieve a phase shift of
$2.2^{\circ} \pm 0.5^{\circ}$ at approximately half linewidth
detuning.
These values are compared to the theoretically predicted values of
Eq. \ref{eq:phaseJJ} expected for a coupling efficiency of 
$G = 13.7 \pm 1.4\,\%$,
found in an independent experiment based on a saturation
measurement\,\cite{alber2016}.
For detunings $-\Gamma/2\le\Delta\le0$ the measured phase shift
shows a systematic deviation from the theoretical model, which
assumes a detuning independent coupling parameter $G$. 
There are three possible reasons for these deviations. Firstly,
saturation effects are neglected for the theoretical curve based on
Eq. \ref{eq:phaseJJ}. However, since the saturation of the
transition was kept low during the measurements ($S<0.1$), the
reduction of the measured phase due to saturation effects is
expected to be less than 2\%. Secondly, the observed phase shift
drops faster than expected when going closer to resonance. A
possible reason for this is that the temperature of the atom
diverges when the detuning approaches zero \cite{chang2014} and
consequently the size of the ion's wave function increases
\cite{eschner2003}. This leads to a stronger averaging of
the experienced electric field by the extent of the ions
wave function \cite{teo2011} entailing a reduction of the coupling
efficiency and therefore also of the measured phase shift.
Lastly, measuring the phase shift via heterodyne
detection not only leads to a phase shift of the close to resonant
part of the two light fields focused on the ion but also to a
non-zero phase shift of the $400\,\textrm{MHz}$ detuned sideband,
acting as a phase reference. At about $400\;\textrm{MHz}$ detuning,
this phase shift of the reference beam can be assumed constant over
the measured data points and takes a value of approximately
$0.13^\circ$ at a coupling efficiency of $G=13.7\%$, leading to an
effective offset of the zero phase value, which is neglected in
Fig.\,\ref{fig:results}.

\section{Concluding remarks}

The phase shift obtained in our experiments is among the largest phase
shifts measured for a coherent beam interacting with a single emitter
in free space so far\,\cite{PhysRevLett_kurtsiefer, Sandoghdar_phase, Blatt_phase}.
Nevertheless it still is far below the maximum possible value
$\Delta\varphi=\pi$ which can be obtained on resonance for
$G>0.5$\,\cite{leuchs2013o, sondermann2013p, zumofen2009}.  
The lower phase shift demonstrated in our experiments is in parts due
to the motion of the ion in the trap and the 
aberrations imprinted by the parabolic mirror, which made it necessary
to focus only from half solid angle.
The latter restriction limits the coupling to
$G\le0.5$\,\cite{golla2012, leuchs2013o}.
But the more severe limitation is the choice of our atomic
species with its reduced scattering cross section.
Even for optimum focusing and cooling the ion to its motional ground
state the imprinted phase shift will never be larger than 30$^\circ$
-- what still appears to be a fairly large value.
Therefore, besides compensating mirror aberrations we aim at repeating
our experiments with $^{174}\text{Yb}^{2+}$\,\cite{heugel_resonant_2016},
which offers the desired $J'=1\leftrightarrow J=0$ transition that
enables the maximum scattering cross section.

\section*{Acknowledgements}
G.L. gratefully acknowledges financial support from the European
Research Council via the Advanced Grant \lq PACART\rq.
We acknowledge the contributions of R.~Maiwald, A.~Golla, S.~Heugel
and M.~Bader in the earlier stages of our experimental endeavors and
useful comments on the manuscript by M.R.~Foreman.


\begin{thebibliography}{34}
\bibitem [{min()}]{minA_note}%
  \BibitemOpen
  \href@noop {} {}\bibinfo {note} {We interprete the effective mode area $A$ as
  the power $P$ of the incident light divided by the intensity $I$ at the
  position of the atom. With the help of Eqs. 11 and 12 from
  Ref.\,\cite{leuchs2013o} one can directly obtain $A=3\lambda^2/(8\pi\cdot G)$
  and thus $A=\sigma_{\text{sc}}/(4G)$, which is minimized in free space by
  $G=1$.}\BibitemShut {Stop}%
\end{thebibliography}
%

\end{document}